\documentclass[rnote]{aa} 
\usepackage{graphicx}
\usepackage{txfonts}
\begin{document}
\title{Stars and brown dwarfs in the $\sigma$~Orionis cluster. III}  
\titlerunning{OSIRIS/GTC spectroscopy in $\sigma$~Orionis}
\subtitle{OSIRIS/GTC low-resolution spectroscopy of variable sources}
\author{J. A. Caballero\inst{1}, A. Cabrera-Lavers\inst{2,3,4}, D. Garc\'{\i}a-\'Alvarez\inst{2,3,4}, S. Pascual\inst{5}}
\institute{
Centro de Astrobiolog\'{\i}a (CSIC-INTA), European Space Astronomy Centre, PO~Box 78, 
28691 Villanueva de la Ca\~nada, Madrid, Spain
\email{caballero@cab.inta-csic.es}
\and
Instituto de Astrof\'{\i}sica de Canarias, Avenida V\'{\i}a L\'actea, 
38205 La Laguna, Tenerife, Spain
\and
Grantecan S.\,A., Centro de Astrof\'{\i}sica de La Palma, Cuesta de San Jos\'e,
38712 Bre\~na Baja, La Palma, Spain 
\and
Departamento de Astrof\'{\i}sica, Universidad de La Laguna, 
38205 La Laguna, Tenerife, Spain
\and
Departamento de Astrof\'{\i}sica y Ciencias de la Atm\'osfera, Facultad de F\'{\i}sica, Universidad Complutense de Madrid, 
28040 Madrid, Spain}
\date{Received 18 May 2012; accepted 13 September 2012}

\abstract
{Although many studies have been performed so far, there are still dozens of low-mass stars and brown dwarfs in the young $\sigma$~Orionis open cluster without detailed spectroscopic characterisation.}
{We look for unknown strong accretors and disc hosts that were undetected in previous surveys.}  
{We collected low-resolution spectroscopy (R $\sim$ 700) of ten low-mass stars and brown dwarfs in $\sigma$~Orionis with OSIRIS at the Gran Telescopio Canarias under very poor weather conditions.
These objects display variability in the optical, infrared, H$\alpha$, and/or X-rays on time scales of hours to years. 
We complemented our spectra with optical and near-/mid-infrared photometry.}     
{For seven targets, we detected lithium in absorption, identified H$\alpha$, the calcium doublet, and forbidden lines in emission, and/or determined spectral types for the first time.
We characterise in detail a faint, T~Tauri-like brown dwarf with an 18\,h-period variability in the optical and a large H$\alpha$ equivalent width of --125$\pm$15\,{\AA}, as well as two M1-type, X-ray-flaring, low-mass stars, one with a warm disc and forbidden emission lines, the other with a previously unknown cold disc with a large inner hole.}  
{New unrevealed strong accretors and disc hosts, even below the substellar limit, await discovery among the list of known $\sigma$~Orionis stars and brown dwarfs that are variable in the optical and have no detailed spectroscopic characterisation yet.}
\maketitle
%

\section{Introduction}
\label{introduction}

   \begin{table*}
      \caption[]{Log of OSIRIS/GTC observations$^a$.} 
         \label{table.log}
     $$ 
         \begin{tabular}{c ll cc ccc cl}
            \hline
            \hline
            \noalign{\smallskip}
	& Name  					& Alternative 				& $\alpha$ 	& $\delta$		& Date of		& $t_{\rm exp}$  		& Airmass			& Seeing			& Trans-		\\ 
	&	  					& name					& (J2000)		& (J2000)		& observation	& [s]  				&  				& [arcsec] 			& parency$^b$\\ %
            \noalign{\smallskip}
            \hline
            \noalign{\smallskip}
	& \object{Mayrit~783254} 		& 2E 1455				& 05 37 54.40	& --02 39 29.8	& 04 Mar 2012	& 3 $\times$ 30				& 1.82		& $\gtrsim$2.5		& Thin		\\ 
	& \object{Mayrit~528005} AB	& [W96] 4771--899			& 05 38 48.03	& --02 27 14.1	& 18 Mar 2012	& 3 $\times$ 45				& 1.86		& $\gtrsim$1.8		& Clear		\\ 
	& \object{Mayrit~609206} 		& V505 Ori				& 05 38 27.25	& --02 45 09.6	& 18 Mar 2012	& 3 $\times$ 45				& 1.48		& $\gtrsim$1.4		& Clear		\\ 
{*}	& \object{Mayrit~957055} 		& [SWW2004] 163			& 05 39 37.30	& --02 26 56.8	& 04 Mar 2012	& 1 $\times$ 600			& 1.37		& $\lesssim$1.5	& Thin		\\ 
{*}	& \object{Mayrit~180277}$^c$	& [W96] rJ053832--0235b		& 05 38 32.84	& --02 35 39.2	& 04 Mar 2012	& 2 $\times$ 300			& 1.20		& $\lesssim$1.4	& Thin		\\ 
{*}	& \object{Mayrit~156353} 		& [W96] rJ053843--0233		& 05 38 43.55	& --02 33 25.4	& 04 Mar 2012	& 2 $\times$ 300			& 1.25		& $\lesssim$1.6	& Thin		\\ 
{*}	& \object{Mayrit~662301} 		& Kiso A--0904 67			& 05 38 06.74	& --02 30 22.8	& 04 Mar 2012	& 2 $\times$ 300			& 1.30		& $\lesssim$1.5	& Thin		\\ 
{*}	& \object{Mayrit~1298302} 	& [SWW2004] 137			& 05 37 31.54	& --02 24 27.0	& 04 Mar 2012	& 1 $\times$ 600			& 1.57		& $\lesssim$2.5	& Thin		\\ 
{*}	& \object{Mayrit~872139}$^d$	& [BZR99] S\,Ori~28			& 05 39 23.19	& --02 46 55.8	& 31 Mar 2012	& 1 $\times$ 900			& 2.01		& 1.5				& Clear		\\ 
{*}	& \object{Mayrit~1196092}$^e$& J054004.5--023642		& 05 37 31.54	& --02 24 27.0	& 12 Mar 2012	& 1 $\times$ 900			& 1.84		& 1.5				& Clear		\\ 
	& \object{UCM0535--0246} 	& 2E 1456				& 05 37 56.31	& --02 45 13.1	& 17 Mar 2012	& 1 $\times$ 1200			& 1.99		& $\gtrsim$2.5		& Clear		\\ 
	& \object{LP 379--51} 			& GJ 3790 (M2.0\,V)			& 13 31 50.57	&  +23 23 20.3	& 16 Mar 2012$^f$	& 2 $\times$ 30				& 1.22		& 1.3				& Thick		\\ 
	& \object{Ross 1022} 			& GJ 3795 (M3.0\,V)			& 13 38 37.05 	&  +25 49 49.7	& 16 Mar 2012$^f$	& 3 $\times$ 30				& 1.19		& 1.2				& Thick		\\ 
	& \object{LP 799--7} 				& GJ 3820 (M4.0\,V)			& 13 59 10.46 	& --19 50 03.5	& 16 Mar 2012$^f$	& 3 $\times$ 60				& 1.51		& 1.2				& Thick		\\ 
	& \object{FN Vir} 				& GJ 493.1 (M4.5\,V)			& 13 00 33.51 	&  +05 41 08.2	& 16 Mar 2012$^f$	& 2 $\times$  60			& 1.20		& 1.3				& Thick		\\ 
	& \object{LP 380--6} 				& GJ 1179 A (M5.5\,V)		& 13 48 13.41 	&  +23 36 48.8	& 16 Mar 2012$^f$	& 2 $\times$ 90				& 1.17		& 1.3 			& Thick		\\ 
	& \object{LP 731--58} 			& GJ 3622 (M6.5\,V)			& 10 48 12.58	& --11 20 08.2	& 22 Mar 2012	& 2 $\times$ 60				& 1.33		& $\gtrsim$1.6		& Thick		\\ 
         \noalign{\smallskip}
            \hline
         \end{tabular}
     $$ 
\begin{list}{}{}
\item[$^{a}$] Asterisks mark our main targets.
\item[$^{b}$] Atmospheric conditions -- 
Clear: clear sky, but with some rare clouds; 
Thin: thin cirrus, inducing absorption up to 0.2\,mag;
Thick: cloudy, inducing absorption over 0.2\,mag.
\item[$^{c}$] Mayrit~180277 was also observed on 06 Mar 2012 at airmass of 1.25 for 900\,s with the volume-phase holographic grating R2500V (R $\sim$ 1400, $\Delta \lambda \sim$ 4400--6000\,{\AA}).
\item[$^{d}$] The spectrum of Mayrit~872139 was flagged as of poor quality (``PQC'') during the quality control due to the high air mass of the observation and the faintness of the target. 
\item[$^{e}$] The full alternative name of Mayrit~1196092 is [BMZ2001] S\,Ori J054004.5--023642. 
\item[$^{f}$] These were the notes in the William Herschel Telescope observing log for the night of 16 Mar 2012: ``strong extinction through clouds most of the night [...]''.
\end{list}
   \end{table*}

The young \object{$\sigma$~Orionis} open cluster ($\tau \sim$ 3\,Ma, $d \sim$ 385\,pc), in the vicinity of the \object{Horsehead Nebula}, is one of the most attractive and most visited regions for night-sky observers.
With its abundant, continuous, and assorted population of early-type, Herbig~Ae/Be, T~Tauri, and FU~Orionis stars, Herbig-Haro objects, brown dwarfs, and planetary-mass objects, $\sigma$~Orionis is also a cornerstone for the study of the initial mass function, disc evolution, X-ray emission, and accretion at all mass domains (Garrison 1967; Wolk 1996; B\'ejar et~al. 1999; Zapatero Osorio et~al. 2000; Walter et~al. 2008; Caballero 2008b and references therein). 
In the past five years, there have been numerous studies in  $\sigma$~Orionis: 
massive spectroscopic analyses of intermediate resolution (Maxted et~al. 2008; Sacco et~al. 2008; Gatti et~al. 2008), 
confirmation and detection of infrared flux excess in isolated planetary-mass objects with discs (Scholz \& Jayawardhana 2008; Luhman et~al. 2008), 
new investigations in X-rays with {\em Chandra} and {\em XMM-Newton} (L\'opez-Santiago \& Caballero 2008; Skinner et~al. 2008; Caballero et~al. 2010a), discoveries of photoevaporated proplyds and cloud overdensities in the cluster centre (Caballero et~al. 2008; Bouy et~al. 2009; Hodapp et~al. 2009), 
or deep photometric surveys for very faint objects (Lodieu et~al. 2009; Bihain et~al. 2009; Pe\~na-Ram\'{\i}rez et~al. 2011; B\'ejar et~al. 2011).
At this point, the reader may think that $\sigma$~Orionis is a drained, barren land where no more outstanding fruits can be gathered.
This is far from reality, considering the recent studies of the central $\sigma$~Ori star itself, ``the most massive binary with an astrometric orbit'', which is actually a triple system of over 40\,$M_\odot$ (Sim\'on-D\'{\i}az et~al. 2011), of \object{Mayrit~264077}, an object at the cluster substellar boundary with the largest modeled total disc mass among a list of 47 targets investigated with {\em Herschel} (Harvey et~al. 2012), and of the VISTA Orion survey, whose findings doubled the number of $\sigma$~Orionis planetary-mass candidates known to date (Pe\~na-Ram\'{\i}rez et~al. 2012). 
This is an impressive result, since there are (at least) four times more objects beyond the deuterium-burning mass limit with spectroscopic characterisation in $\sigma$~Orionis than in the whole remaining sky.  

One way of gathering those ``outstanding fruits'' in $\sigma$~Orionis is to spectroscopically investigate poorly known targets towards the cluster with chances to show some spectrophotometric peculiarity.
The choice of these targets can be random or based on the presence of unusual features, such as strong X-flare activity or high-amplitude variability in the optical.
Here we used OSIRIS at the Gran Telescopio Canarias for obtaining low-resolution spectroscopy of eleven variable sources towards $\sigma$~Orionis, with a variety of magnitudes ($J \approx$ 9.3--15.3\,mag) and origins of variability.

\section{Observations and analysis}
\label{section.obsanalysis}

  \begin{table*}
    \caption[]{Known data of sources in $\sigma$~Orionis$^{a}$.} 
      \label{table.knowndata}
        $$ 
	\begin{tabular}{l cc c c c c}
\hline
\hline
\noalign{\smallskip}
Name  		        	& Phot.			& X-ray			& Spectral			& EW(H$\alpha$)			& EW(Li~{\sc i})				& $V_r$ 					\\
	  	        		& variability  		& variability		& type  			& [\AA] 					& [m\AA] 					& [km\,s$^{-1}$]				\\
\noalign{\smallskip}
\hline
\noalign{\smallskip}
Mayrit~783254 	& ...				& Flare (Ca09)		& K0 (Wo96)		& +0.521 (Wo96)			& +0.417 (Wo96)			& ...						\\ %
Mayrit~528005\,AB   & Mid-term (Ca10b)	& ...				& K7.0 (ZO02)$^b$	& --4.3$\pm$0.3 (Ca06)$^b$	& +470$\pm$20 (Ca06)$^b$	& +30.7$\pm$0.4 (GH08)		\\ %
Mayrit~609206 	& Mid-term (Fe60)$^c$& ...			& K7.0e (ZO02) 	& --25.11$\pm$0.72 (Sa08)$^c$& +431$\pm$11 (Sa08)		& +30.00$\pm$0.49 (Sa08)	\\ %
Mayrit~957055 	& ...				& Long-term (Ca10a)& ...				& ...						& ...						& ...						\\ %
Mayrit~180277 	& Mid-term (CH10)	& Flare (Ca10a)	& K8.5: (Sa08)		& --1.46$\pm$0.17 (Sa08)	& +576$\pm$6 (Sa08)		& +30.12$\pm$0.17 (Sa08) 	\\ %
Mayrit~156353 	& ...				& Flare (Fr06)$^d$	& M1.0: (Sa08)		& --3.59$\pm$0.29 (Sa08)	& +620$\pm$23 (Sa08)		& +31.63$\pm$0.33 (Sa08) 	\\ %
Mayrit~662301 	& ...				& Flare (Fr06)		& ...				& Variable (Wi89)$^e$		& ...						& ...						\\ %
Mayrit~1298302 	& ...				& Flare (LC08)		& ...				& Emission? (Ca06)$^f$		& Absorption? (Ca06)$^f$	& ...						\\ 
Mayrit~872139 	& Pulsation? (Ca04)$^g$	& ...			& ...				& ...						& +660$\pm$90 (Ke05)		& +24.84$\pm$0.53 (Ma08) 	\\ %
Mayrit~1196092 	& Short-term (Ca04)$^h$	& ...			& ...				& Broad (Ke05)				& +390: (Ke05)				& +31.45$\pm$0.47 (Ma08) 	\\ %
\noalign{\smallskip}
\hline
	\end{tabular}
     	$$ 
	\begin{list}{}{}
	\item[$^{a}$] Reference acronyms in parenthesis --
	Fe60: Fedorovich (1960);	
	ZO02: Zapatero Osorio et~al. (2002);
	Wi89: Wiramihardja et~al. (1989);
	Wo96: Wolk (1996);
	Ca04: Caballero et~al. (2004);
	Ke05: Kenyon et~al. (2005);
	Ca06: Caballero (2006);
	Fr06: Franciosini et~al. (2006);
	GH08: Gonz\'alez Hern\'andez et~al. (2008);
	LC08: L\'opez-Santiago \& Caballero (2008);
	Ma08: Maxted et~al. (2008);
	Sa08: Sacco et~al. (2008); 
	Ca09: Caballero et~al. (2009);
	Ca10a: Caballero et~al. (2010a); 
	CH10: Cody \& Hillenbrand (2010);
	Ca10b: Caballero et~al. (2010b). 
	\item[$^{b}$] For Mayrit~528005\,AB, Zapatero Osorio et~al. (2002) had previously measured EW(H$\alpha$) = --3.1$\pm$0.5\,{\AA} and EW(Li~{\sc i}) = +480$\pm$70\,m{\AA}, while Wolk (1996) had estimated a K3 spectral type.
	\item[$^{c}$] For Mayrit~609206, Cody \& Hillenbrand (2010, 2011) also found aperiodic photometric variability in the optical and near-infrared, while Zapatero Osorio et~al. (2002) and Caballero (2006) had previously measured EW(H$\alpha$) = --53.5$\pm$9.0 and --32$\pm$3 \,{\AA}, respectively.
	\item[$^{d}$] For Mayrit~156353, Caballero et~al. (2009, 2010a) also found X-ray flaring activity.
	\item[$^{e}$] Wiramihardja et~al. (1989) tabulated different estimated H$\alpha$ emission intensities of Mayrit~662301 at three epochs: medium, very weak, and absent or doubtful.
	\item[$^{f}$] The AF2+WYFFOS/4.2\,m William Herschel Telescope spectrum of Mayrit~1298302 in Caballero (2006) was very noisy. 
	\item[$^{g}$] For Mayrit~872139, Cody \& Hillenbrand (2010) did not re-detect variability in their data; any signal would be below 0.004\,mag at periods shorter than 8\,h and $\sim$0.01\,mag for longer timescales of up to two weeks.
	\item[$^{h}$] For Mayrit~1196092, Cody \& Hillenbrand (2010) determined a period $P$ = 0.76$\pm$0.01\,d with $A_I$ = 0.027$\pm$0.010\,mag.
	\end{list}
  \end{table*}

We carried out low-resolution spectroscopy with the Optical System for Imaging and Low Resolution Integrated Spectroscopy (OSIRIS) tunable imager and spectrograph (Cepa et~al. 2003; Cepa 2010) at the 10.4\,m Gran Telescopio Canarias (GTC), located at the Observatorio Roque de los Muchachos in La Palma, Canary Islands, Spain. 
The heart of OSIRIS is a mosaic of two 4k\,$\times$\,2k e2v CCD44--82 detectors that gives an unvignetted field of view of 7.8\,$\times$\,7.8\,arcmin$^{2}$ with a plate scale of 0.127\,arcsec\,pix$^{-1}$. 
However, to increase the signal-to-noise ratio of our observations, we chose the standard operation mode of the instrument, which is a 2\,$\times$\,2-binning mode with a readout speed of 100\,kHz. 

All spectra were obtained with the OSIRIS R1000B grism. 
We used the 1.23\,arcsec-width slit, oriented at the parallactic angle to minimise losses due to atmospheric dispersion. 
The resulting resolution, measured on arc lines, was R $\sim$ 700 in the approximate 3500--8000\,{\AA} spectral range. 

Observations were performed in service mode within the GTC ``filler'' programme GTC55--12A on different nights in March 2012. 
Night conditions were quite variable, covering a wide range of different weather conditions. 
The aim of this filler programme was to obtain, within the GTC nightly operation schedule, high-quality spectra of {\em variable} sources with no spectral type determination that are relatively bright for a 10\,m-class telescope (with magnitudes of up to $V \sim$ 19\,mag) in poor weather conditions, such as dust presence, sky brightness, dense cirrus coverage, or poor seeing (of over 1.4\,arcsec). 
As an example, some of the 16 Mar 2012 spectra were taken while the rest of La Palma telescopes were closed.
This circumstance made the quality of our spectra be very different from one night to another without direct relation to the target brightness.

The names, coordinates, and main observing parameters of the 17 observed targets are listed in Table~\ref{table.log}.
Seven of them, marked with asterisks, made up our main target sample. 
They were low-mass star and brown dwarf cluster members and member candidates without spectral-type determination and with variability found  in the optical, infrared, H$\alpha$, and/or X-rays at short- (hours), mid- (days), and/or long-term (months and years) time scales.
Two of them had spectral types estimated from $R-I$ colours, while another two had lithium and radial velocity measurements only. 
We also targeted three well-known $\sigma$~Orionis K-type variable stars for comparison purposes.
We provide the known variability and spectroscopic data with suitable references of the ten stars and brown dwarfs in Table~\ref{table.knowndata}.
We took our filler programme beyond the limit with the observation of the two faintest brown dwarfs, which have $V$ magnitudes of over 21\,mag and were observed at very low airmasses.

Furthermore, we also observed six field M-dwarf comparison stars, shown at the bottom of Table~\ref{table.log}. 
Since we found some discordances with spectral typing as tabulated by the Simbad database, we used only the types from Hawley et~al. (1996 -- see a justification by Jim\'enez-Esteban et~al. 2012).
The six M dwarfs cover the spectral range from M2.0\,V to M6.5\,V, where we expected our faintest targets in $\sigma$~Orionis to lie.
Finally, an active galaxy in the cluster background also satisfied our programme requirements: to be variable (in X-rays) and to have no spectroscopic characterisation (in the optical).
The strong X-ray source 2E~1456 (UCM0535--0246 -- L\'opez-Santiago \& Caballero 2008; Caballero et~al. 2009) turned out to be a Seyfert~1 galaxy with wide ($\sim$750\,km\,s$^{-1}$) forbidden and very wide ($\sim$2500\,km\,s$^{-1}$) Balmer lines in emission at a cosmological redshift $z_{\rm sp}$ = 0.10960$\pm$0.00006.

\begin{figure}
\centering
\includegraphics[width=0.49\textwidth]{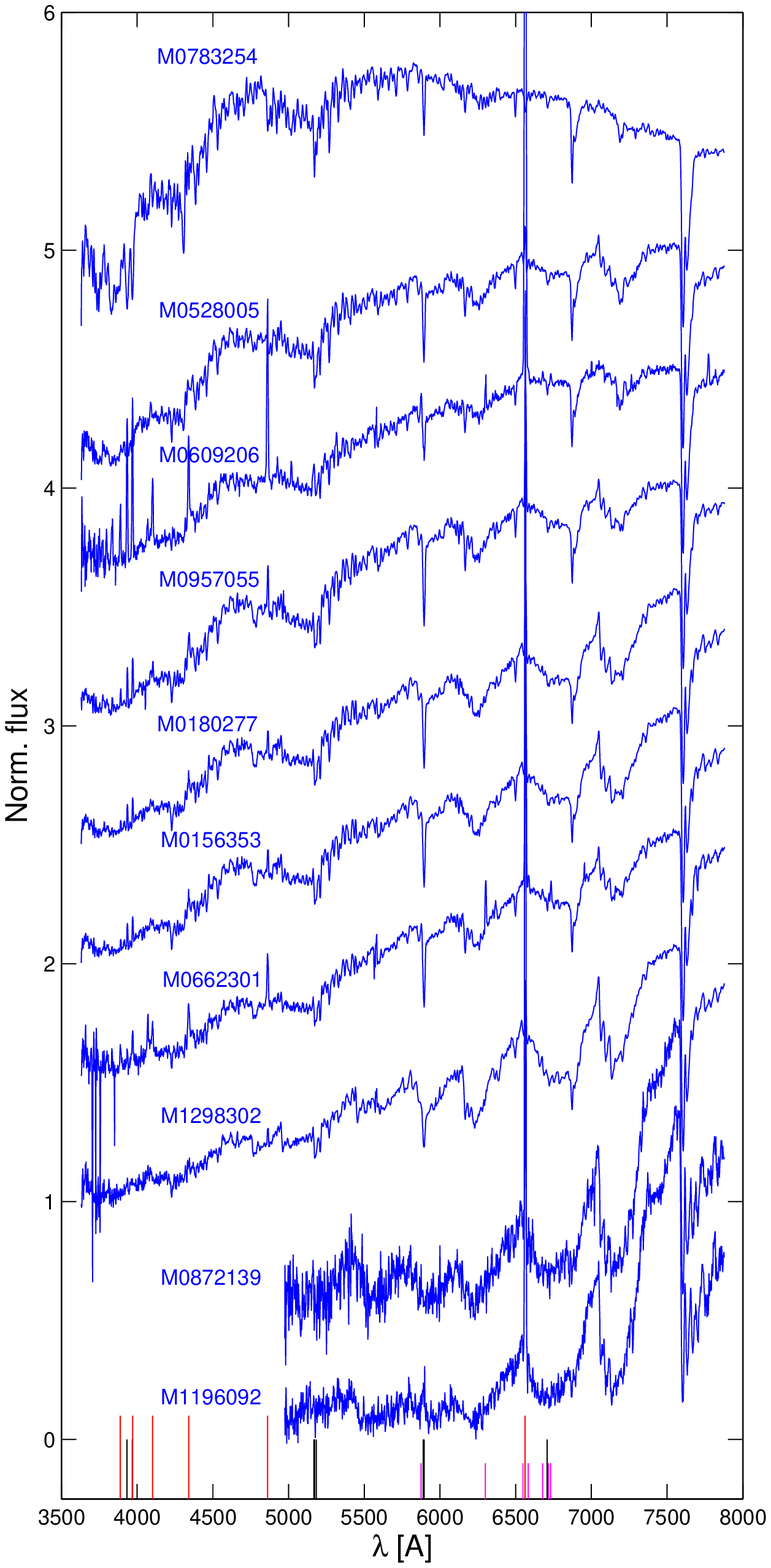}
\caption{OSIRIS/GTC low-resolution spectra of the ten $\sigma$~Orionis stars and brown dwarfs.
Names are labelled.
The noisy region of the brown dwarf Mayrit~872139 and~1196092 spectra bluewards of 5000\,{\AA} are not plotted.
In the bottom, vertical lines indicate key spectral signposts: the hydrogen Balmer series (very long [red] lines), calcium, sodium, magnesium, and lithium in absorption (long [black] lines), and helium and forbidden oxygen, nitrogen, and sulfur in emission (short [magenta] lines).} %
\label{figure.spectraOri}
\end{figure}

We used standard tasks within the IRAF software collection for the spectra reduction.
Images were initially bias-subtracted and flat-field corrected using lamp flats from the GTC instrument calibration module. 
The two-dimensional spectra were then wavelength-calibrated using Xe+Ne+HgAr lamps. 
Next, sky background was subtracted and a one-dimension spectrum was optimally extracted with an extraction aperture that varied depending on the seeing of the corresponding exposure. 
The instrument-response correction was applied using the spectrophotometric standard white dwarfs BD+52~913 (G\,191--B2b), Feige\,34, and Ross\,640 from the list of Oke (\cite{oke74}, \cite{oke90}) observed on the same nights as the scientific data with the 2.5\,arcsec slit. 
When more than one spectrum was available for a target, they were averaged to obtain a mean spectrum. 
The normalised, fully-reduced spectra of the ten young stars and brown dwarfs are shown in Fig.~\ref{figure.spectraOri}.

We measured full-widths at half maximum and equivalent widths (EWs) of spectral lines of the targets with the IRAF task {\tt onedspec.splot}. 
We focused on singlets and doublets typical of T~Tauri stars: the Balmer series (from H$\alpha$~$\lambda$6562.8\,{\AA} to H$\zeta$~$\lambda$3889.1\,{\AA}),  the calcium doublet (Ca~H\,\&\,K $\lambda\lambda$3933.7,\,3968.5\,{\AA}), and forbidden lines of [N~{\sc ii}]~$\lambda\lambda$6548.0,\,6583.4\,{\AA}, [S~{\sc ii}]~$\lambda\lambda$6716.4,\,6730.8\,{\AA} and [O~{\sc i}]~$\lambda$6300.3\,{\AA} in emission, and Li~{\sc i}~$\lambda$6707.8\,{\AA} in absorption.
The low spectral resolution prevented us from measuring the broadening of the H$\alpha$ line.

Spectral types of the ten young stars and brown dwarfs were derived by comparing their spectra with those of well-known field K and M dwarfs (Gray \& Corbally 2009) and our comparison stars in Table~\ref{table.log}.  
Uncertainties of our spectral type determinations are of one subtype redwards of K5.

\section{Results}
\label{section.results}

  \begin{table*}
    \caption[]{New spectroscopic data of stars and brown dwarfs in $\sigma$~Orionis.} 
      \label{table.results}
        $$ 
	\begin{tabular}{l c cccccc c}
\hline
\hline
\noalign{\smallskip}
Name  		        		& Sp.	        		& EW(Ca K)		& EW(Ca H)	        	& EW(H$\gamma$)	& EW(H$\beta$) 	& EW(H$\alpha$)		& EW(Li~{\sc i})		& SED	\\
	  	        			& type          		& [\AA] 			& [\AA] 	        		& [\AA] 	        		& [\AA] 			& [\AA] 				& [\AA] 			& class$^{a}$	\\
\noalign{\smallskip}
\hline
\noalign{\smallskip}
Mayrit~783254 	        	& K0:	        		& +13$\pm$2		& +9$\pm$2	        	& +0.7$\pm$0.2	& +0.8$\pm$0.6  	&   +0.8$\pm$0.2  		& +0.62$\pm$0.15	& III		\\ %
Mayrit~528005\,AB   	& K7.0$\pm$1.0  	& +4.5$\pm$0.2	& +4.5$\pm$0.4	& $<$ +0.4		& $<$ +0.2		&  --1.9$\pm$0.2		& +0.6$\pm$0.2	& II		\\ %
Mayrit~609206 	        	& K7.0$\pm$1.0e  	& --26$\pm$3		& --26$\pm$3	        	& --9$\pm$2	        	& --15.0$\pm$1.0	&   --46$\pm$2			& +0.45$\pm$0.15	& II		\\ 
Mayrit~957055 	        	& M0.0$\pm$1.0  	& --13$\pm$2		& --12$\pm$2	        	& --3.5$\pm$1.5	& --3.5$\pm$0.5	&  --5.5$\pm$0.3		& +0.65$\pm$0.15	& III		\\ %
Mayrit~180277 	        	& M1.0$\pm$1.0  	& --7$\pm$2		& --8$\pm$2	        	& --2.5$\pm$1.5        	& --3.0$\pm$1.0	&  --4.4$\pm$0.2		& +0.60$\pm$0.15	& III:		\\ %
Mayrit~156353 	        	& M1.0$\pm$1.0  	& --10$\pm$2		& --9.5$\pm$1.0        	& --2.0$\pm$0.5        	& --2.5$\pm$0.2	&  --3.9$\pm$0.2		& +0.55$\pm$0.15	& II		\\ %
Mayrit~662301 	        	& M1.0$\pm$1.0e  	& --5.5$\pm$1.5	& --5.5$\pm$1.0        	& --9$\pm$2	        	& --7.0$\pm$1.0	& --11.8$\pm$0.9		& +0.55$\pm$0.10	& II		\\ 
Mayrit~1298302 		& M3.0$\pm$0.5  	& $>$ --2.0		& $>$ --2.0		& $>$ --1.0		& --2.6$\pm$0.2	&  --3.5$\pm$0.3		& $<$ +0.2		& III:		\\ 
Mayrit~872139 	        	& M5.5$\pm$1.0  	& ...				& ...				& ...				& ...				&   --13$\pm$3			& $<$ +1.0		& III		\\ 
Mayrit~1196092 		& M6.0$\pm$1.0e 	& ...				& ...				& ...				& ...				&  --125$\pm$15		& $<$ +1.0		& II		\\ 
\noalign{\smallskip}
\hline
	\end{tabular}
     	$$ 
	\begin{list}{}{}
	\item[$^{a}$] Spectral-energy-distribution class:
	``II'': with a disc;
	``III'': without a disc;
	``III:'': probably without a disc.
	\end{list}
 \end{table*}

We provide in Table~\ref{table.results} the equivalent widths of key lines in our spectra of ten stars and brown dwarfs in $\sigma$~Orionis. 
Although the Li~{\sc i} line is blended at this resolution with faint lines of Fe~{\sc i} at 6703.6, 6705.1, and 6710.3\,{\AA} and of CN (e.g., Ghezzi et~al. 2009), its detection is clear in the spectra of the seven brightest stars. 
Thus, we measure for the first time lithium in Mayrit~957055 and 662301 and confirm their extreme youth.

The H$\alpha$ line is in emission in all young objects but one.
Three of them, indicated by ``e'' in the second column in Table~\ref{table.results}, satisfy the empirical criterion for classifying T~Tauri stars and substellar analogues (Barrado y Navascu\'es \& Mart\'{\i}n 2003).
The only young object not displaying H$\alpha$ in emission is the K0 star Mayrit~783254, which also shows the rest of Balmer lines in absorption and has a larger uncertainty in the estimation of its spectral type (indicated in Table~\ref{table.results} with a colon).
We quantify the H$\alpha$ emission for the first time of five young objects, of which two are accretors.
They are the low-mass star Mayrit~662301 and the brown dwarf Mayrit~1196092.  
As a result, this substellar object displays all key signposts typical of T~Tauri stars: optical variability, X-ray emission, infrared-flux excess, and, from now on, H$\alpha$ in strong emission, which falls in the highest quartile among members in the Ori~OB1 association.
The T~Tauri nature of the third accreting object Mayrit~609206 (V505~Ori), was already known.
With the four EW(H$\alpha$) measures collected in Tables~\ref{table.knowndata} and~\ref{table.results}, Mayrit~609206 becomes one of the few $\sigma$~Orionis stars whose {variable} H$\alpha$ emission has been monitored for over a decade. 

As expected, we find a correlation between the equivalent widths of the H$\alpha$ line, the rest of Balmer lines, and the Ca H\,\&\,K doublet (at least for the seven brightest targets; e.g., Mart\'{\i}nez-Arn\'aiz et~al. 2011).
To our knowledge, this is the first time that the calcium doublet is investigated down to the M spectral type in $\sigma$~Orionis (Muzerolle et~al. 2003 tabulated equivalent widths of H$\gamma$ and the calcium infrared triplet at 8500--8660\,{\AA}, but not of the Ca H\,\&\,K doublet).
However, the strongest calcium emission is found in a K7 star, Mayrit~609206.
Previously, Gatti et~al. (2008) had found Pa$\beta$, Pa$\gamma$, and He~{\sc i} $\lambda$1.083\,$\mu$m in emission in the near-infrared spectrum of this accreting star.
Not surprisingly, the OSIRIS/GTC spectrum of Mayrit~609206 also displays [N~{\sc ii}], [S~{\sc ii}], and [O~{\sc i}] in emission, as Zapatero Osorio et~al. (2002) and Caballero (2006) had measured previously.
The [O~{\sc i}] emission supports the hypothesis that the K7 star is the host of a small Herbig-Haro object (cf. Caballero 2006).
The other target displaying [N~{\sc ii}], [S~{\sc ii}], and [O~{\sc i}] in emission in our set is the second-brightest accretor, Mayrit~662301.
We speculate that if not for the poor signal-to-noise ratio we would have also been able to identify these forbidden lines in the spectrum of the T~Tauri-like brown dwarf Mayrit~1196092. 

The spectral types derived by us match previous determinations based on real spectra (the three K-type comparison stars) and optical colours (the two stars in Sacco et~al. 2008).
We derive spectral types for the first time for the remaining five objects, in the range M0--6. 
Despite the numerous spectroscopic studies in the last decade in $\sigma$~Orionis, only three works apart from ours have derived early M spectral types (M0--4) of low-mass stars: Wolk (1996), Zapatero Osorio et~al. (2002), and Caballero et~al. (2008).
This is because of the short wavelength coverage of the instrument setups used (e.g., Andrews et~al. 2004; McGovern et~al. 2004; Kenyon et~al. 2005; Caballero 2006; Jeffries et~al. 2006; Gatti et~al. 2008; Gonz\'alez Hern\'andez et~al. 2008; Maxted et~al. 2008; Sacco et~al. 2008) or a bias towards late spectral types (e.g., B\'ejar et~al. 1999; Mart\'{\i}n et~al. 2001; Barrado y Navascu\'es et~al. 2001, 2003; Scholz \& Eisl\"offel 2004; Caballero et~al. 2006).
There are still a few dozen early M stars in $\sigma$~Orionis waiting for spectral type determination.

We revisited the spectral energy distribution classification of the ten stars and brown dwarfs in the last column of Table~\ref{table.results}.
For that, we collected photometry in the red optical ($R$ UCAC3: Zacharias et~al. 2010; $r'$ SDSS DR8: Aihara et~al. 2011), near-infrared ($JHK_{\rm s}$ 2MASS: Skrutskie et~al. 2006), and mid-infrared ($W1 W2 W3 W4$ {\em WISE} at 3.35, 4.6, 11.6, and 22.1\,$\mu$m: Wright et~al. 2010).
Three stars and brown dwarfs were not detected at the $W4$ (and $W3$) passband: Mayrit~783254, 957055, and 872139.
Of the remaining objects, four were already known to host a disc (i.e., are of class~II): Mayrit~528005\,AB, 609206, 662301, and 1196092 (Hern\'andez et al. 2007; Caballero et al. 2007).
The M1 star Mayrit~156353 displays a clear flux excess redwards of 10\,$\mu$m ($W2-W3$ = 0.71 $\pm$ 0.05\,mag, $W3-W4$ = 3.94 $\pm$ 0.08\,mag) and, therefore, we classify it as a disc-host star. 
Interestingly, it did not show any flux excess in the IRAC/{\em Spitzer} bands and did not show up at all in the MIPS/{\em Spitzer} data (Hern\'andez et~al. 2007). 
This may be because its disc is transitional and has a large inner hole.
The $W4$ magnitudes of Mayrit~180277 and 1298302 have an uncertainty one order of magnitude larger than the rest of the studied sources and are probably erroneous, so previous classifications of the two stars as class~III are likely correct.
The latter star displayed a large X-ray flare during the {\em XMM-Newton} observations in L\'opez-Santiago \& Caballero (2008) and has a low PPMXL proper motion, of $\mu_\alpha \cos{\delta}$ = --8.9$\pm$3.8\,mas\,a$^{-1}$, $\mu_\delta$ = --0.1$\pm$3.8\,mas\,a$^{-1}$ (Roeser et~al. 2010).
However, Mayrit~1298302 is located at over 21.6\,arcmin from the cluster centre, where the spatial density of cluster members drops sharply and the probability of finding interlopers increases (Caballero 2008a; Lodieu et~ al. 2009; B\'ejar et~al. 2011; Pe\~na-Ram\'{\i}rez et~al. 2012), and the two available spectra (in Caballero 2006 and in this work) do not show lithium with confidence.
As a result, we considered Mayrit~1298302 as a photometric cluster member candidate only.

\section{Discussion}
\label{section.discussion}

\begin{figure}
\centering
\includegraphics[width=0.49\textwidth]{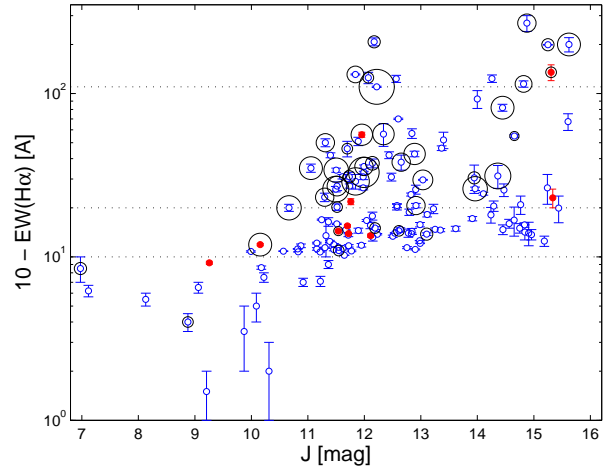}
\caption{H$\alpha$ equivalent width vs. $J$-band magnitude diagram in $\sigma$~Orionis.
Radii of large (black) circles are proportional to the peak-to-peak amplitude of known photometric variability in the optical.
Filled (red) objects are objects investigated here.
Dotted horizontal lines denote EW(H$\alpha$) = 0, --10, and --100\,{\AA}, from bottom to top.
The cluster stellar-substellar boundary lies at about $J$ = 14.5\,mag.} %
\label{figure.HalphaJvariability}
\end{figure}

One would expect that variability is directly linked to other features typical in T~Tauri stars and substellar analogues (e.g., accretion; Bertout 1989).
Nevertheless, the sources of variability in our work are very heterogeneous, both in wavelength (X-rays, H$\alpha$, optical, and infrared) and in time sampling (to minutes, through hours and days, to years).
To investigate the correlation between accretion and {\em optical} variability in $\sigma$~Orionis, we first exhaustively searched the literature and compiled the 146 cluster stars and brown dwarfs brighter than $J$ = 16\,mag with published measurement of EW(H$\alpha$) (B\'ejar et~al. 1999; Zapatero Osorio et~al. 2002; Barrado y Navascu\'es et~al. 2003; Andrews et~al. 2004; Scholz \& Eisl\"offel 2004; Caballero 2006; Caballero et~al. 2006, 2008; Sacco et~al. 2008; Oksala et~al. 2012).
Of these, 20 have at least two EW(H$\alpha$) measurements.
The earliest and latest spectral types of the investigated targets are B2Vp (\object{$\sigma$~Ori~E}) and $\sim$M6 (\object{Mayrit~358154}), respectively.
In Fig.~\ref{figure.HalphaJvariability}, we plot the EW(H$\alpha$) as a function of $J$ magnitude, which is a rough proxy of mass.
The strongest accretors are located in the upper part of the diagram, the most massive ones to the left.

Among the 146 cluster members, 42 have reported peak-to-peak amplitudes of variability in the optical (e.g., Fedorovich 1960; North 1984; Scholz \& Eisl\"offel 2004; Caballero et~al. 2004, 2006, 2010b; Cody \& Hillenbrand 2010).
They are plotted with large circles in Fig.~\ref{figure.HalphaJvariability}.
In general, the most variable sources have the largest EW(H$\alpha$)s in their magnitude bin.
In particular, of the 11 stars and brown dwarfs with EW(H$\alpha$) $\le$ --100\,{\AA}, nine were known to vary.
One of them is the brown dwarf Mayrit~1196092, whose EW(H$\alpha$) is measured here for the first time.
The other two strong accretors for which photometric variability has not been reported (yet) are Mayrit~91024 and 36263, possibly due to their closeness to the cluster centre (monitorings have tended to avoid the glare of the central OB star system). 
Numerous stars and brown dwarfs with EW(H$\alpha$)s between --50 and --100\,{\AA} remain as potential variables, too.
On the whole, stars with X-ray flaring activity only do not show large amplitudes of optical variability or strong accretion because they are not tidally locked to a disc (fast rotation enhaces magnetic fields -- Feigelson et~al. 1993; Neuh\"auser et~al. 1995; Preibisch \& Zinnecker 2002; Telleschi et~al. 2007).
Not even Mayrit~783254, with its violent long-lived flare detected by {\em ROSAT}, displays the EW(H$\alpha$) found in, e.g., Mayrit~609296, 662301, and 1196092.

To summarise, a programme devoted to obtain low-resolution spectroscopy of young stars and brown dwarfs that are known to vary in the {\em optical} is an efficient way to discover new strong accretors and disc hosts that were undetected in previous surveys.

\begin{acknowledgements}

We thank the anonymous referee for his/her helpful and careful report.
This article is based on observations made with the Gran Telescopio Canarias operated on the island of La Palma by the Instituto de Astrof\'{\i}sica de Canarias in the Spanish Observatorio de El Roque de Los Muchachos.
This research made use of the SIMBAD, operated at Centre de Donn\'ees astronomiques de Strasbourg, France, and the NASA's Astrophysics Data System. 
JAC is an {\em investigador Ram\'on y Cajal} at the Consejo Superior de Investigaciones Cient\'{\i}ficas.
Financial support was provided by the Spanish Ministerio de Ciencia e Innovaci\'on and Ministerio de Econom\'{\i}a y Competitividad under grants
AyA2009-10368 and		
AyA2011-30147-C03-03. 		

\end{acknowledgements}

\end{document}